\documentclass[preprint,12pt]{elsarticle}
\biboptions{numbers,sort&compress} 

\usepackage[english]{babel}
\usepackage{CJKutf8}
\usepackage{booktabs}
\usepackage{longtable}
\usepackage{amssymb}
\usepackage{setspace}
\usepackage{graphicx}
\usepackage{array}
\usepackage{amsmath}
\usepackage{caption}
\usepackage{multirow}
\usepackage[table]{xcolor}
\usepackage[colorlinks=true, allcolors=blue]{hyperref}

\journal{Physica A: Statistical Mechanics and its Applications}

\begin{document}
\begin{CJK}{UTF8}{gbsn}

\begin{frontmatter}

\title{Impacts of Data Splitting Strategies on Parameterized Link Prediction Algorithms}

\author[uestc]{Xinshan Jiao}
\author[UoW]{Yuxin Luo}
\author[uestc]{Yilin Bi}
\author[uestc]{Tao Zhou\corref{cor1}}

\cortext[cor1]{Corresponding author: Tao Zhou}
\ead{zhutou@ustc.edu}

\address[uestc]{CompleX Lab, School of Computer Science and Engineering, University of Electronic Science and Technology of China, Chengdu 610054, China}

\address[UoW]{Department of Computer Science, University of Warwick, Coventry CV4 7AL, United Kingdom}

\begin{highlights}
\item Revealing information leakage in traditional parameter training for link prediction.
\item Proposing a method to measure the impact of information leakage.
\item Demonstrating its significant impact on evaluation results via large-scale experiments.
\end{highlights}


\begin{abstract}
Link prediction is a fundamental problem in network science, aiming to infer potential or missing links based on observed network structures. With the increasing adoption of parameterized models, the rigor of evaluation protocols has become critically important. However, a previously common practice of using the test set during hyperparameter tuning has led to human-induced information leakage, thereby inflating the reported model performance. To address this issue, this study introduces a novel evaluation metric, Loss Ratio, which quantitatively measures the extent of performance overestimation. We conduct large-scale experiments on 60 real-world networks across six domains. The results demonstrate that the information leakage leads to an average overestimation of about 3.6\%, with the bias reaching over 15\% for specific algorithms. Meanwhile, heuristic and random-walk-based methods exhibit greater robustness and stability. The analysis uncovers a pervasive information leakage issue in link prediction evaluation and underscores the necessity of adopting standardized data splitting strategies to enable fair and reproducible benchmarking of link prediction models.
\end{abstract}

\end{frontmatter}

\section{Introduction}
Link prediction is a fundamental problem in network science, aiming to infer potential or future connections between pairs of nodes based on observed topological information~\cite{lu2011,wang2015,martinez2016,kumar2020,divakaran2020,zhou2021,arrar2024}. The practical relevance of link prediction spans a wide range of domains. In social networks, it aids in identifying latent social relationships; in biological networks, it facilitates the discovery of previously unverified protein-protein interactions; and in recommender systems, it enhances the identification of user preferences and improves recommendation quality~\cite{adamic2003,kovacs2019,lu2012}. Consequently, link prediction has become an essential analytical tool for understanding, modeling, and leveraging the complex relational structures. Over the past few decades, link prediction techniques have undergone substantial evolution. The field has evolved from simple similarity indices based on local structural patterns~\cite{katz1953,adamic2003,leicht2006,liben2007,fouss2007,zhou2009,liu2010,cannistraci2013}, through parameterized models with enhanced expressive power~\cite{clauset2008,guimera2009,soundarajan2012,lu2015}, and more recently toward deep graph representation learning methods~\cite{kunegis2009, bordes2013,perozzi2014,yang2015,trouillon2016,grover2016,kipf2017,hamilton2017,velickovic2018,zhang2018,xu2019,ahn2021,chamberlain2022}. Research efforts continue to advance link prediction in both theoretical foundations and practical applications. As real-world networks continue to expand in scale, heterogeneity, temporal variability, and diversity of structure formation, link prediction algorithms have generally become more complex, with larger model sizes, higher-dimensional parameter spaces, and more context-dependent~\cite{zhou2021, bi2025domain}. Consequently, the reliability and fairness of performance evaluation procedures have become critically important. Evaluation outcomes not only determine the credibility of empirical comparisons across different methods but also directly influence the methodological trajectory and pace of development in this field. If the evaluation process contains biases or implicit assumptions, algorithmic performance may be inaccurately estimated, leading to misleading conclusions and adverse implications for subsequent research and applications.

In the evaluation of parameterized link prediction algorithms, a critical yet often overlooked issue is the improper use of data splitting strategies. Conventional studies predominantly adopt the two-set split, i.e., the train–test split, in which the link set is divided into a training set and a test set. The model is first trained on the training subset, and hyperparameter tuning is subsequently conducted using the test set. This strategy has been widely adopted due to the simplicity of implementation~\cite{kohavi1995,arlot2010}. However, the two-set split introduces a fundamental methodological flaw because the model receives feedback from the test set during hyperparameter tuning, which compromises the independence of evaluation. Specifically, in link prediction tasks, researchers typically explore a range of hyperparameter combinations and compute evaluation metrics on the test set, selecting the configuration that yields the best performance (e.g., the highest AUC value). Although this process may seem reasonable, it effectively allows the model to iteratively adjust its hyperparameters based on feedback from the test set~\cite{kaufman2012}, thereby inducing a dependency of the learned parameters on the test data. Although the test links are not explicitly used during training, this practice nevertheless violates the basic principle of evaluation independence. The model indirectly acquires structural information from the test set during hyperparameter tuning, resulting in \textbf{information leakage}. Here, classical data leakage typically refers to sample-level overlap between the training and test data. In contrast, the information leakage discussed in this work specifically denotes an evaluation-induced leakage mechanism, arising from test-dependent hyperparameter selection and manifesting as systematic evaluation bias. In effect, the test set no longer represents truly unseen data; the situation is analogous to having prior access to exam questions before taking a test, granting the model an unwarranted advantage during evaluation. For parameterized algorithms, such leakage leads to inflated test performance, the model appears to perform well because it has been implicitly tuned toward patterns specific to the test set. In contrast, non-parametric algorithms do not involve hyperparameter tuning and thus cannot exploit feedback from the test set, creating an inherent inequity between the two classes of methods~\cite{dietterich1998,cawley2010}. The informational asymmetry compromises the comparability of algorithms and prevents experimental outcomes from accurately reflecting model performance in truly unknown environments, thereby undermining the reliability and reproducibility of link prediction research.

In the field of machine learning, a fundamental principle for preventing information leakage is the strict separation of parameter selection from performance evaluation~\cite{cawley2010,varma2006,forman2010}. Accordingly, when applied to the link prediction task, this principle naturally adopts the three-set splitting strategy, i.e., the train-validation-test split, in which the link set is partitioned into mutually exclusive training set, validation set, and test set. The training set is used for model fitting, the validation set is employed for hyperparameter tuning or early stopping, and the test set is reserved exclusively for the final evaluation after the model is fully specified. By design, the three-set split prevents premature exposure to test data and provides a more faithful assessment of link prediction performance.

However, due to historical inertia and experimental convenience, the two-set split continues to be widely adopted in link prediction research~\cite{lu2011,wang2015,martinez2016,kumar2020,divakaran2020,zhou2021,arrar2024}. In many early studies, test set information was mistakenly used during the hyperparameter tuning stage, enabling parameterized algorithms to gain unfair performance advantages and leading to misleading conclusions regarding their true generalization ability on unseen data~\cite{lu2011}. Although the three-set split provides an effective solution to this issue, its application in the link prediction domain remains insufficiently widespread. Furthermore, quantitative studies that systematically compare the evaluation discrepancies between the two-set and three-set splits across different algorithmic paradigms and network categories are still lacking. 

Existing literature has recognized that data splitting strategies can significantly affect the reliability of model evaluation, yet most discussions are fragmented and limited to specific domains. For example, Roberts \textit{et al.}~\cite{roberts2017} pointed out that conventional random splitting in ecological data breaks sample independence, consequently underestimating prediction error, and proposed a structured cross-validation strategy to improve the fairness of generalization assessment. Breit \textit{et al.}~\cite{breit2020} introduced the OpenBioLink benchmarking framework to reveal potential information leakage and unfair comparisons caused by generic data partitioning strategies in link prediction evaluation, and provided a unified and reproducible evaluation protocol. Similarly, Varoquaux and Cheplygina~\cite{varoquaux2022} further observed that similar information leakage issues are prevalent in medical machine learning and often lead to inflated evaluation results. Overall, although these studies reveal that improper evaluation procedures may cause biases, they primarily focus on domain-specific scenarios and do not offer a unified quantitative framework applicable across algorithms and network categories.

To analyze the gap between the two-set and the three-set splits from information leakage, we will introduce the metric called \textbf{Loss Ratio} to quantify the performance discrepancy and robustness variation induced by the two-set split, compared to the three-set split, in parameterized link prediction tasks. Thereby, we establish a more equitable and reproducible evaluation standard for link prediction. Many representative algorithms spanning multiple methodological paradigms, including heuristic methods, random-walk-based methods,  similarity-transfer methods, and deep learning methods are selected for analysis, including Katz~\cite{katz1953}, LHN-II~\cite{leicht2006}, Local Path (LP)~\cite{zhou2009}, Local Random Walk (LRW)~\cite{liu2010}, Superposed Random Walk (SRW)~\cite{liu2010}, Random Walk with Restart (RWR)~\cite{fouss2007}, TSAA~\cite{sun2009}, TSCN~\cite{sun2009}, DeepWalk~\cite{perozzi2014}, Graph Convolutional Network (GCN)~\cite{kipf2017} and Variational Graph Normalized Autoencoder (VGNAE)~\cite{ahn2021}. This comprehensive comparison across disparate algorithms and networks not only reveals the potential impact of information leakage on the fairness of evaluation but also provides empirical evidence for establishing a more rigorous evaluation framework in the domain of link prediction.

\section{Methods}

To characterize the biases introduced by different data splitting strategies, we propose a unified framework that directly quantifies the performance differences caused by test information leakage. This framework compares the performance differences between the two strategies on a common evaluation set, ensuring the fairness. Given a network $G(\mathcal{V},\mathcal{E})$, this framework is designed according to two principle: (1) nested partitioning with a consistent test set, and (2) a unified partition ratio. Specifically, both the two-set and three-set splits are evaluated on an identical test set, ensuring that any observed performance difference can be attributed solely to the partitioning strategy. If we denote $E_T$, $E_V$, and $E_P$ as the training, validation, and test link sets $(\mathcal{E}=E_T \cup E_V \cup E_P)$, respectively, and $E_T’$ as the training set used in the two-set split. Under the two-set split, the training set $E_T'$ is first extracted from the link set $\mathcal{E}$ and the remaining links are designated as the unified test set $E_P$. Subsequently, we further divide $E_T'$ into the training set $E_T$ and the validation set $E_V$, as required by the three-set split, such that $E_T' = E_T \cup E_V,\quad E_T \cap E_V=\varnothing,\quad E_T \cap E_P=\varnothing,\quad E_V \cap E_P=\varnothing$. In the precedure of link splitting, we preserve the connectivity of the training network $G(\mathcal{V},E_T)$. More specifically, we will resplit the links if the current splitting results in a disconnected training network.

\begin{figure}[h!]
    \centering
    \includegraphics[width=1\textwidth]{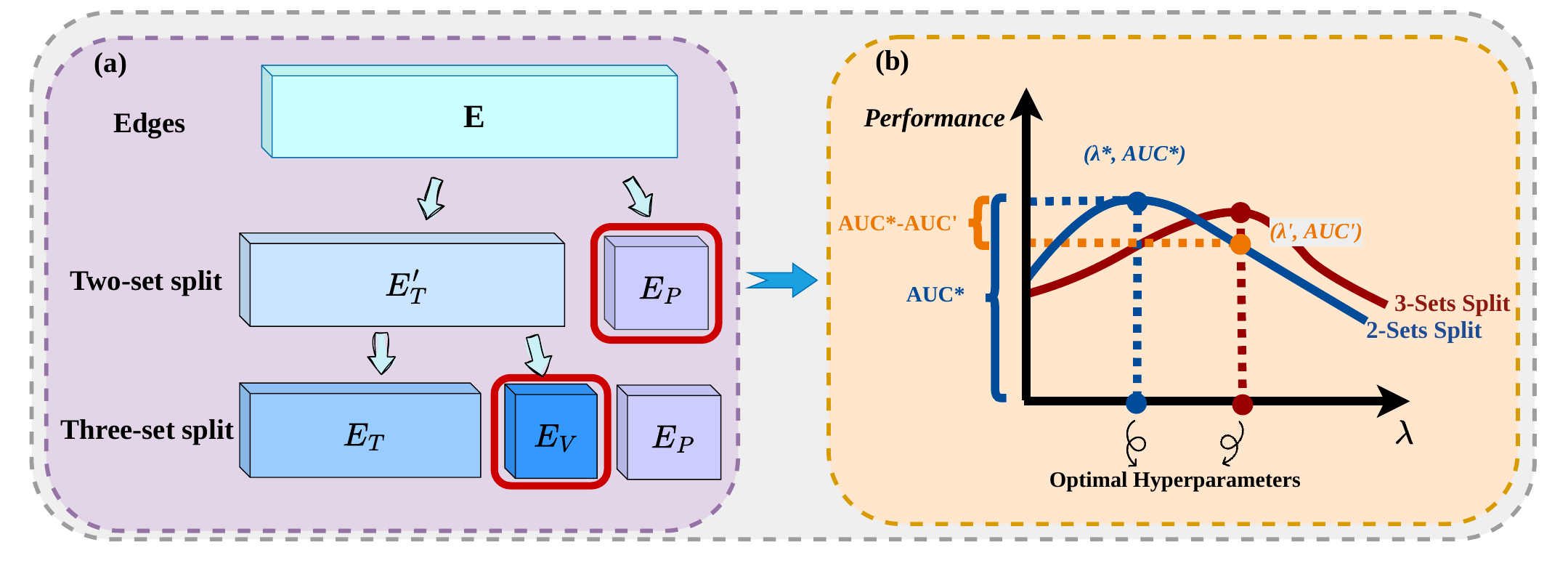}
    \caption{
    The two-set and three-set split diagram.
    \textbf{(a)} The edge set $\mathcal{E}$ of the original network is divided under two strategies. 
    In the two-set split, $\mathcal{E}$ is partitioned into the training set $E_T'$ and test set $E_P$. 
    In the three-set split, $E_T'$ is further divided into the training set $E_T$ and the validation set $E_V$, while both strategies share the same test set $E_P$ through a nested splitting scheme.
    The red boxes indicate where the optimal hyperparameters are selected in each strategy. 
    \textbf{(b)} Illustration of AUC–$\lambda$ performance curves. The blue curve represents performance evaluated on $E_P$ (corresponding to the two-set split), while the red curve corresponds to performance evaluated on $E_V$ (corresponding to the three-set split). The point $(\lambda^\ast,\mathrm{AUC}^\ast)$ denotes the optimal hyperparameter and corresponding performance obtained from $E_P$, and $(\lambda',\mathrm{AUC}')$ represents the optimal hyperparameter selected on the $E_V$ and its mapped performance on $E_P$. The difference between these two reflects the performance degradation caused by hyperparameter transfer, thereby quantifying the impact of information leakage.
    }
    \label{23_Diagram}
\end{figure}

To further control the randomness introduced by data splitting, a proportional parameter $\rho \in (0,1)$ is introduced to ensure consistent relative sizes across the subsets. In the three-set split, the proportions of the training, validation, and test sets are set as $(1-\rho)^2 : (\rho - \rho^2) : \rho$, which guarantees $
|E_T| : |E_V| = (|E_T| + |E_V|) : |E_P| = (1-\rho) : \rho$.
Accordingly, for the two-set split, the partition satisfies
$|E_T'| : |E_P| = (1-\rho) : \rho, \quad |E_T'| = |E_T| + |E_V|$. Through these two principles, both $E_T$ and $E_V$ are strictly derived from $E_T'$, and their relative sizes are uniformly controlled by the parameter $\rho$, which maintains consistency across different networks and sampling strategies. By eliminating discrepancies in sample proportions and reducing the impact of random partitioning, this nested scheme allows performance differences to accurately reflect the intrinsic effects of the partitioning strategy itself.

Under the two-set split, the training set $E_T'$ and the test set $E_P$ are jointly used for both model training and hyperparameter tuning. In contrast, under the three-set split, the training set $E_T$ is used to learn the model parameters, the validation set $E_V$ is employed for hyperparameter selection, and the test set $E_P$ is strictly reserved for final evaluation only after the model has been fully determined. To enable a direct comparison between the two strategies, the optimal hyperparameter $\lambda'$ obtained by the validation set in the three-set split is projected onto the test set of the two-set split and compared with the performance corresponding to the hyperparameter $\lambda^*$ that is obtained directly by the test set (with information leakage). The core procedure is illustrated in Fig.~\ref{23_Diagram}. Importantly, the two-set split is used solely as a contrastive reference for quantifying
evaluation bias induced by information leakage, rather than as a recommended or acceptable
evaluation procedure.

Given that the primary objective of this work is to analyze the evaluation bias induced by different data splitting strategies, rather than to optimize top-ranked prediction performance in a specific application scenario, we adopt the Area Under the ROC Curve (AUC) as the main evaluation metric, following a series of studies on link prediction metrics~\cite{zhou2023, wan2025, bi2024, jiao2024}. In the context of link prediction, AUC represents the probability that a randomly selected existing link is assigned a higher prediction score than a randomly selected non-existing link, thereby providing a threshold-free and global assessment of ranking performance. This property makes AUC a stable and neutral basis for comparing evaluation biases across different algorithms and network structures. Then, the \textit{Loss Ratio} metric is defined as
\[
  \mathcal{L} = \frac{|\mathrm{AUC}^\ast - \mathrm{AUC}'|}{\mathrm{AUC}^\ast}, \quad \mathcal{L} \in [0,1],
\]
where $\mathrm{AUC}^\ast$ denotes the optimal performance achieved under the two-set split, and $\mathrm{AUC}'$ denotes the performance obtained by applying the optimal hyperparameter selected under the three-set split to the test set. A smaller value of $\mathcal{L}$ indicates that the two splitting strategies yield similar results, implying strong robustness of the algorithm. Conversely, a larger value of $\mathcal{L}$ suggests that the information leakage, introduced by the two-set split, leads to significant overestimation of test performance and an inaccurate reflection of the model’s true generalization capability on unseen data.

A variety of representative link prediction models covering four methodological paradigms are systematically compared in this study. These algorithms encompass both traditional methods based on local and global structural features, as well as recent embedding-based models, thus providing strong representativeness and diversity.

Heuristic methods mainly rely on the topological information of the network.  Katz~\cite{katz1953} measures the global connectivity between nodes by summing over all possible paths with an exponential decay factor, thereby capturing long-range structural dependencies.  LHN-II~\cite{leicht2006} can be regarded as a normalized variant of the Katz index, recursively defining node similarity based on the similarity of their neighbors, i.e., if two nodes have similar neighbors, they are also considered similar. LP~\cite{zhou2009} extends the Common Neighbor by incorporating three-order path information, balancing local accuracy and computational efficiency.  

Random-walk-based methods estimate node similarity by simulating information propagation over the network. LRW~\cite{liu2010} performs a finite-step random walk starting from a given node and defines similarity according to the probability of visiting another node within a fixed number of steps.  SRW~\cite{liu2010} aggregates results from multiple walk lengths to emphasize nearby nodes and suppress noise from distant ones.  RWR~\cite{fouss2007} measures global structural proximity based on the expected hitting time in a random walk process, a higher visiting probability between nodes $i$ and $j$ implies greater structural similarity.  

Similarity-transfer methods are based on the assumption that similarity between nodes is transitive.  TSAA and TSCN~\cite{sun2009} extend the Adamic--Adar and Common Neighbor indices, respectively, by introducing a decay factor to aggregate multi-step similarities, thereby capturing indirect structural associations across the network.  

Deep learning methods automatically capture dependencies between nodes through graph representation learning.  
DeepWalk~\cite{perozzi2014} introduces the Skip-Gram model from natural language processing into network representation learning. It performs multiple random walks on the graph to generate node sequences and trains a neural language model to learn low-dimensional embeddings, where structurally similar nodes are mapped close to each other in the embedding space.  
Graph Neural Networks (GNNs)~\cite{kipf2017,velickovic2018,hamilton2017} aggregate neighborhood features through a message-passing mechanism to capture higher-order structural information. In this study, GCN is implemented within the Graph Autoencoder (GAE) framework~\cite{kipf2017}, where graph convolution enables smooth feature propagation and structural integration.  
VGNAE~\cite{ahn2021} extends the Variational Graph Autoencoder (VGAE)~\cite{kipf2017} by introducing normalization constraints and a KL-divergence regularization term. By modeling the uncertainty of latent node representations, VGNAE enhances robustness against noise and alleviates overfitting.  

\section{Data Description}
This study includes 60 real-world networks~\cite{kunegis2013,rossi2015}, covering six categories: Animal Social (13), Biological (11), Informational (8), Social (10), Technological (8), and Transportation (10). The structural statistics of those networks are reported in Table~\ref{DetailofNetwork}, where $N=|\mathcal{V}|$ denotes the number of nodes, $M=|\mathcal{E}|$ represents the number of links, $\langle k\rangle$ is the average degree, and $r$ denotes the link density (defined as the ratio of actual links to the theoretical maximum number of links). Overall, those networks vary substantially in both size and density, spanning several orders of magnitude. The diversity ranges from sparse biological and engineering systems to relatively dense social and informational networks, demonstrating that the datasets provide broad coverage and strong representativeness.

\small                 
\setlength\tabcolsep{1pt}  
\renewcommand\arraystretch{0.62538}
\setlength\LTleft{0pt}       
\setlength\LTright{0pt}  

\begin{longtable}{@{}>{\centering\arraybackslash}m{2.9cm}
                      >{\centering\arraybackslash}m{5.8cm}
                      >{\centering\arraybackslash}m{0.9cm}
                      >{\centering\arraybackslash}m{0.9cm}
                      >{\centering\arraybackslash}m{1.2cm}
                      >{\centering\arraybackslash}m{1.5cm}@{}}
  \caption{Structural statistics of networks.\label{DetailofNetwork}}\\
  \toprule
  Domain & Name &$N$&$M$&$\langle k\rangle$&$r$\\
  \midrule
  \endfirsthead

  \toprule
  Domain & Name &$N$&$M$&$\langle k\rangle$&$r$\\
  \midrule
  \endhead

  \bottomrule
  \multicolumn{6}{@{}r}{}\\
  \endfoot

  \bottomrule
  \endlastfoot
  \multirow{13}{*}{Animal Social}
    & mammalia-voles-bhp-trapping-28 &  153 &  189 &  2.47 & 0.0163 \\
    & mammalia-voles-plj-trapping-27 &  125 &  229 &  3.66 & 0.0295 \\
    & mammalia-voles-bhp-trapping-22 &  115 &  239 &  4.16 & 0.0365 \\
    & mammalia-voles-bhp-trapping-23 &  128 &  253 &  3.95 & 0.0311 \\
    & reptilia-tortoise-network-fi-2008 &  283 &  418 &  2.95 & 0.0105 \\
    & reptilia-tortoise-network-fi-2011 &  437 &  479 &  2.19 & 0.0050 \\
    & mammalia-voles-plj-trapping-25 &  168 &  503 &  5.99 & 0.0359 \\
    & mammalia-dolphin-floridatravel &  188 & 1032 & 10.98 & 0.0587 \\
    & mammalia-dolphin-florida-forage &  190 & 1134 & 11.94 & 0.0632 \\
    & reptilia-tortoise-network-fi   &  787 & 1197 &  3.04 & 0.0039 \\
    & aves-weaver-social             &  445 & 1332 &  5.99 & 0.0135 \\
    & mammalia-dolphin-florida-overall &  291 & 3182 & 21.87 & 0.0754 \\
    & insecta-ant-colony1-day11      &  113 & 3454 & 61.13 & 0.5458 \\
  \midrule
  \multirow{12}{*}{Biological}
    & bn-mouse-visual-cortex-2       &  193 &  214 &  2.22 & 0.0116 \\
    & maayan-pdzbase                 &  212 &  242 &  2.28 & 0.0108 \\
    & gene-fusion                    &  291 &  279 &  1.92 & 0.0066 \\
    & bio-DM-LC                      &  658 & 1129 &  3.43 & 0.0052 \\
    & bio-CE-LC                      & 1387 & 1648 &  2.38 & 0.0017 \\
    & bio-yeast                      & 1458 & 1948 &  2.67 & 0.0018 \\
    & arenas-meta                    &  453 & 2025 &  8.94 & 0.0198 \\
    & bio-celegans                   &  453 & 2025 &  8.94 & 0.0198 \\
    & Celegans-w                     &  297 & 2148 & 14.46 & 0.0489 \\
    & moreno-propro-propro           & 1846 & 2203 &  2.39 & 0.0013 \\
    & foodweb-little-rock            &  183 & 2434 & 26.60 & 0.1462 \\
  \midrule
  \multirow{8}{*}{Informational}
    & GD06-theory                    &  101 &  190 &  3.76 & 0.0376 \\
    & fs-adjnoun-adj-copperfield     &  112 &  425 &  7.59 & 0.0684 \\
    & webkb-texas-link1              &  285 &  493 &  3.46 & 0.0122 \\
    & webkb-wisconsin-link1          &  297 &  565 &  3.80 & 0.0129 \\
    & email-enron-only               &  143 &  623 &  8.71 & 0.0614 \\
    & webkb-cornell-link1            &  349 &  696 &  3.99 & 0.0115 \\
    & wiki-talk-ht                   &  446 &  758 &  3.40 & 0.0076 \\
    & webkb-washington-link1         &  433 &  954 &  4.41 & 0.0102 \\
  \midrule
  \multirow{10}{*}{Social}
    & moreno-highschool              &   70 &  274 &  7.83 & 0.1135 \\
    & ia-enron-only                  &  143 &  623 &  8.71 & 0.0614 \\
    & moreno-innovation              &  241 &  923 &  7.66 & 0.0319 \\
    & librec-filmtrust-trust         &  874 & 1309 &  3.00 & 0.0034 \\
    & moreno-oz-oz                   &  217 & 1839 & 16.95 & 0.0785 \\
    & arenas-jazz                    &  198 & 2742 & 27.70 & 0.1406 \\
    & dimacs10-netscience            & 1461 & 2742 &  3.75 & 0.0026 \\
    & comp2012Net                    & 2175 & 4541 &  4.18 & 0.0019 \\
    & comp2014Net                    & 2401 & 5930 &  4.94 & 0.0021 \\
    & petster-friendships-hamster-uniq & 1858 & 12534 & 13.49 & 0.0073 \\
  \midrule
  \multirow{8}{*}{Technological}
    & 81-ISCAS89-s1208               &  131 &  190 &  2.90 & 0.0223 \\
    & 83-ISCAS89-s1298               &  156 &  264 &  3.38 & 0.0218 \\
    & 86-ISCAS89-s1382               &  183 &  313 &  3.42 & 0.0188 \\
    & 88-ISCAS89-s1400               &  187 &  327 &  3.50 & 0.0188 \\
    & 91-ISCAS89-s1444               &  210 &  363 &  3.46 & 0.0165 \\
    & 89-ISCAS89-s1420               &  269 &  400 &  2.97 & 0.0111 \\
    & 466-5728dc4b461efe0c7a675f37   &  362 &  617 &  3.41 & 0.0094 \\
    & 77-ISCAS89-s1238               &  567 & 1070 &  3.77 & 0.0067 \\
  \midrule
  \multirow{10}{*}{Transportation}
    & urban-brasilia                 &  179 &  230 &  2.57 & 0.0144 \\
    & urban-washington               &  192 &  302 &  3.15 & 0.0165 \\
    & 410-59488c124ed8f1bb6022a8df   &  224 &  376 &  3.36 & 0.0151 \\
    & urban-new-york                 &  248 &  418 &  3.37 & 0.0136 \\
    & 411-59488c124ed8f1bb6022a8de   &  397 &  644 &  3.24 & 0.0082 \\
    & urban-vienna                   &  467 &  691 &  2.96 & 0.0064 \\
    & euroroad                       & 1174 & 1417 &  2.41 & 0.0021 \\
    & ATC                            & 1226 & 2408 &  3.93 & 0.0032 \\
    & opsahl-powergrid               & 4941 & 6594 &  2.67 & 0.0005 \\
    & as20000102                     & 6474 & 12572 &  3.88 & 0.0006 \\
\end{longtable}

\section{Results}

Table~\ref{tab:lossratio} reports the average value of $\mathcal{L}$ for each algorithm within each network category under $\rho = 0.2$, quantifying the degree of performance overestimation induced by the two-set split relative to the three-set split. Overall, the evaluations based on the two-set split lead to an average overestimation of approximately 3.6\%, indicating that the two-set evaluation paradigm systematically inflates the perceived prediction accuracy of parameterized models. 

Notably, the average $\mathcal{L}$ values of different algorithms vary significantly, reflecting substantial differences in their improper gains from information leakage. Specifically, heuristic and random-walk-based algorithms exhibit more robust performance, with most $\mathcal{L}$ values falling below 3\%. This suggests that the performance degradation caused by hyperparameter transfer is relatively minor for these methods. Because these algorithms predominantly rely on local topological features or simplified stochastic processes and involve a limited number of hyperparameters, they are generally less sensitive to the training sample size and data partitioning strategy, thereby maintaining consistent performance across different network structures.

In contrast, deep learning algorithms exhibit substantially higher values of $\mathcal{L}$ across multiple network categories, indicating that their evaluation outcomes are highly sensitive to the choice of data partitioning strategy. The sensitivity may be attributed to two main factors. First, deep models typically contain a large number of trainable parameters and exhibit strong dependence on the size of the training set; when transitioning from the two-set split to the three-set split, the reduced training set size leads to a pronounced decline in performance. Second, deep learning algorithms involve numerous hyperparameters and complex search spaces, resulting in unstable behavior across different networks and amplifying evaluation bias. These factors collectively result in substantially lower robustness of deep models compared to traditional algorithms when the data splitting strategy is altered.

\begin{table}[htbp]
\centering
\renewcommand{\arraystretch}{0.8}
\caption{Average $\mathcal{L}$ across algorithms and network categories under $\rho=0.2$. 
The bottom row (\textbf{Net Avg.}) and the rightmost column (\textbf{Algo Avg.}) represent averages over network categories and algorithms, respectively. The first row lists the abbreviations of different network domains: \textbf{Soc} for Social networks, \textbf{Ani} for Animal Social networks, \textbf{Trans} for Transportation networks, \textbf{Tech} for Technological networks, \textbf{Bio} for Biological networks, and \textbf{Info} for Informational networks.
}
\label{tab:lossratio}
\resizebox{\textwidth}{!}{%
\begin{tabular}{ccccccc!{\vrule width 1pt}c}
\toprule
 & \textbf{Soc} & \textbf{Ani} & \textbf{Trans} & \textbf{Tech} & \textbf{Bio} & \textbf{Info} & \textbf{Algo Avg. } \\
\midrule
\textbf{LP} & \cellcolor[HTML]{73B2D7}\textcolor{black}{0.12\%} & \cellcolor[HTML]{C3D9EE}\textcolor{black}{0.01\%} & \cellcolor[HTML]{85BCDB}\textcolor{black}{0.08\%} & \cellcolor[HTML]{4F9BCB}\textcolor{black}{0.30\%} & \cellcolor[HTML]{2D7DBB}\textcolor{white}{0.94\%} & \cellcolor[HTML]{1C6BB0}\textcolor{white}{1.81\%} & \cellcolor[HTML]{FEE4D8}\textcolor{black}{0.54\%} \\
\textbf{Katz} & \cellcolor[HTML]{EDF4FB}\textcolor{black}{0.00\%} & \cellcolor[HTML]{66AAD4}\textcolor{black}{0.16\%} & \cellcolor[HTML]{4191C5}\textcolor{white}{0.47\%} & \cellcolor[HTML]{2C7CBB}\textcolor{white}{0.99\%} & \cellcolor[HTML]{105CA4}\textcolor{white}{3.13\%} & \cellcolor[HTML]{1461A8}\textcolor{white}{2.57\%} & \cellcolor[HTML]{FCA689}\textcolor{black}{1.22\%} \\
\textbf{LHNII} & \cellcolor[HTML]{519CCC}\textcolor{black}{0.29\%} & \cellcolor[HTML]{539DCC}\textcolor{black}{0.28\%} & \cellcolor[HTML]{2D7DBB}\textcolor{white}{0.93\%} & \cellcolor[HTML]{1562A9}\textcolor{white}{2.48\%} & \cellcolor[HTML]{1B6AAF}\textcolor{white}{1.86\%} & \cellcolor[HTML]{09539D}\textcolor{white}{4.33\%} & \cellcolor[HTML]{FB8A6A}\textcolor{black}{1.70\%} \\
\textbf{SRW} & \cellcolor[HTML]{509BCB}\textcolor{black}{0.30\%} & \cellcolor[HTML]{63A9D3}\textcolor{black}{0.17\%} & \cellcolor[HTML]{3888C0}\textcolor{white}{0.64\%} & \cellcolor[HTML]{1562A9}\textcolor{white}{2.47\%} & \cellcolor[HTML]{1764AB}\textcolor{white}{2.34\%} & \cellcolor[HTML]{084F99}\textcolor{white}{5.04\%} & \cellcolor[HTML]{FB8363}\textcolor{black}{1.83\%} \\
\textbf{LRW} & \cellcolor[HTML]{4E9ACA}\textcolor{black}{0.32\%} & \cellcolor[HTML]{3D8DC3}\textcolor{white}{0.53\%} & \cellcolor[HTML]{2575B7}\textcolor{white}{1.24\%} & \cellcolor[HTML]{1A69AE}\textcolor{white}{1.98\%} & \cellcolor[HTML]{0C57A0}\textcolor{white}{3.78\%} & \cellcolor[HTML]{08519C}\textcolor{white}{4.63\%} & \cellcolor[HTML]{FB7858}\textcolor{black}{2.08\%} \\
\textbf{RWR} & \cellcolor[HTML]{2C7CBB}\textcolor{white}{0.97\%} & \cellcolor[HTML]{3E8EC4}\textcolor{white}{0.52\%} & \cellcolor[HTML]{4191C5}\textcolor{white}{0.46\%} & \cellcolor[HTML]{1360A7}\textcolor{white}{2.70\%} & \cellcolor[HTML]{09539D}\textcolor{white}{4.38\%} & \cellcolor[HTML]{09539D}\textcolor{white}{4.40\%} & \cellcolor[HTML]{FB7252}\textcolor{black}{2.24\%} \\
\textbf{DeepWalk} & \cellcolor[HTML]{1765AB}\textcolor{white}{2.25\%} & \cellcolor[HTML]{0C56A0}\textcolor{white}{3.88\%} & \cellcolor[HTML]{0C56A0}\textcolor{white}{3.96\%} & \cellcolor[HTML]{084A92}\textcolor{white}{5.77\%} & \cellcolor[HTML]{084E97}\textcolor{white}{5.26\%} & \cellcolor[HTML]{084991}\textcolor{white}{5.98\%} & \cellcolor[HTML]{E53127}\textcolor{white}{4.52\%} \\
\textbf{VGNAE} & \cellcolor[HTML]{1C6BB0}\textcolor{white}{1.81\%} & \cellcolor[HTML]{115DA5}\textcolor{white}{2.96\%} & \cellcolor[HTML]{084991}\textcolor{white}{6.00\%} & \cellcolor[HTML]{084489}\textcolor{white}{7.09\%} & \cellcolor[HTML]{08478E}\textcolor{white}{6.33\%} & \cellcolor[HTML]{084991}\textcolor{white}{6.03\%} & \cellcolor[HTML]{DD2924}\textcolor{white}{5.04\%} \\
\textbf{GNN} & \cellcolor[HTML]{1A69AE}\textcolor{white}{1.96\%} & \cellcolor[HTML]{125EA6}\textcolor{white}{2.89\%} & \cellcolor[HTML]{08529C}\textcolor{white}{4.54\%} & \cellcolor[HTML]{084489}\textcolor{white}{7.47\%} & \cellcolor[HTML]{084489}\textcolor{white}{7.09\%} & \cellcolor[HTML]{083B7B}\textcolor{white}{10.16\%} & \cellcolor[HTML]{D31F20}\textcolor{white}{5.69\%} \\
\textbf{TSCN} & \cellcolor[HTML]{1765AB}\textcolor{white}{2.26\%} & \cellcolor[HTML]{0F5AA3}\textcolor{white}{3.41\%} & \cellcolor[HTML]{084B94}\textcolor{white}{5.54\%} & \cellcolor[HTML]{083D7E}\textcolor{white}{9.41\%} & \cellcolor[HTML]{083A7A}\textcolor{white}{10.28\%} & \cellcolor[HTML]{083775}\textcolor{white}{11.67\%} & \cellcolor[HTML]{C1151B}\textcolor{white}{7.09\%} \\
\textbf{TSAA} & \cellcolor[HTML]{2575B7}\textcolor{white}{1.23\%} & \cellcolor[HTML]{0E59A2}\textcolor{white}{3.54\%} & \cellcolor[HTML]{084991}\textcolor{white}{5.93\%} & \cellcolor[HTML]{084489}\textcolor{white}{7.21\%} & \cellcolor[HTML]{083471}\textcolor{white}{12.73\%} & \cellcolor[HTML]{08306B}\textcolor{white}{15.60\%} & \cellcolor[HTML]{BB1419}\textcolor{white}{7.71\%} \\
\midrule
\textbf{Net Avg.} & \cellcolor[HTML]{E8F6E4}\textcolor{black}{1.05\%} & \cellcolor[HTML]{BEE5B7}\textcolor{black}{1.67\%} & \cellcolor[HTML]{81CA81}\textcolor{black}{2.71\%} & \cellcolor[HTML]{3CA659}\textcolor{white}{4.35\%} & \cellcolor[HTML]{2A934B}\textcolor{white}{5.28\%} & \cellcolor[HTML]{157E3A}\textcolor{white}{6.56\%} & \cellcolor{gray!25}3.60\% \\
\bottomrule
\end{tabular}
}
\end{table}

It is noteworthy that the similarity-transfer algorithms TSAA and TSCN, although not deep learning models in a strict sense, also demonstrate high values of $\mathcal{L}$. These methods rely on multi-step propagation over similarity matrices to capture global structural information and often exhibit strong representational capability in sparse networks. However, their performance is highly sensitive to the propagation parameter $\lambda$, increasing the numerical instability and the risk of overfitting. Under the two-set split, such sensitivity is exacerbated by information leakage, leading to inflated performance estimates; under the three-set split, the validation mechanism imposes stricter constraints on hyperparameter selection, causing the performance to drop more substantially, which ultimately results in a higher Loss Ratio. The elevated Loss Ratio observed for these algorithms highlights the potential risk of overestimation under the two-set split and demonstrates that even non-deep models may yield overly optimistic evaluation results if the test set is improperly used during tuning.

\begin{figure}[h!]
\centering
\includegraphics[width=0.8\textwidth]{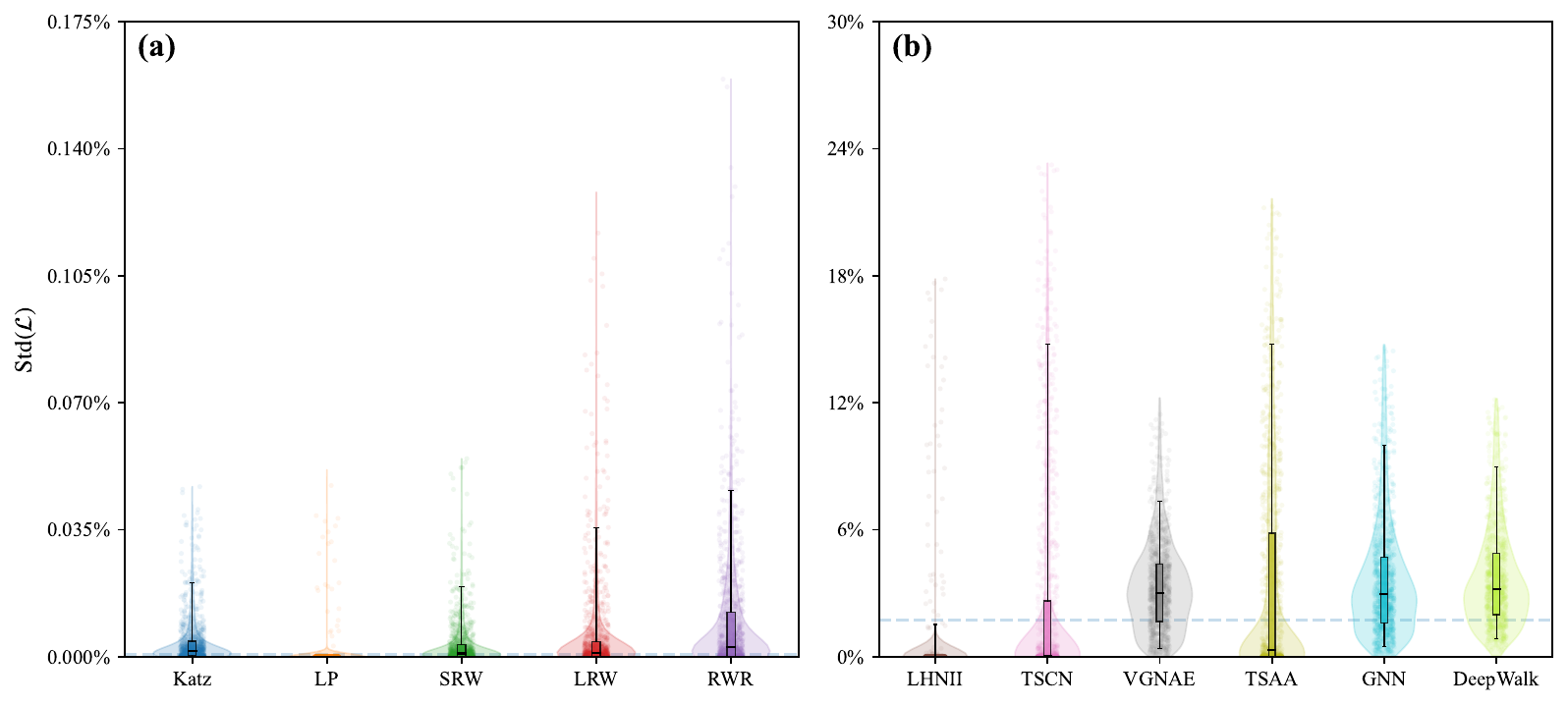}
\caption{Robustness of the loss ratio $\mathcal{L}$ with respect to random data partitioning.
The figure summarizes the distribution of the standard deviation $\mathrm{Std}(\mathcal{L})$ for each algorithm across all settings.
For each setting, $\mathrm{Std}(\mathcal{L})$ is computed from multiple independent repetitions of the data splitting procedure, thereby quantifying the variability induced by partition randomness.
Lower values indicate stronger statistical robustness.
Boxplots report the median and interquartile range, while individual scatter points correspond to specific combinations of network and $\rho$.
}
\label{std_lossratio}
\end{figure}

From the perspective of network categories, Informational and Biological networks exhibit the highest average values of $\mathcal{L}$, reaching 6.56\% and 5.28\%, respectively, indicating that these types of networks are the most sensitive to changes in data splitting strategy. In contrast, Social networks generally show lower $\mathcal{L}$ values. Taken together, the results demonstrate a clear trend that the greater the model complexity and degree of parameterization, the stronger its dependence on the data splitting strategy, and the more likely the two-set split is to produce significantly overestimated performance. This finding substantiates the central claim of this study, namely that unified, reproducible, and fair evaluation standards are essential for parameterized link prediction tasks.

We further examine the statistical robustness of the loss ratio $\mathcal{L}$ with respect to the randomness of data partitioning.
Specifically, for each setting (i.e., algorithm + network + $\rho$), we repeat the data splitting procedure multiple times and compute the resulting standard deviation $\mathrm{Std}(\mathcal{L})$.
Figure~\ref{std_lossratio} shows the distribution of $\mathrm{Std}(\mathcal{L})$ across all settings. We observe that the variability of $\mathcal{L}$ is consistently low for the majority of algorithms. In particular, classical heuristic and random-walk-based methods exhibit near-zero standard deviations across almost all settings, indicating strong robustness to random partitioning. Learning-based methods display moderately higher variability, yet the fluctuations are still well bounded and do not affect the qualitative conclusions. Overall, these results demonstrate that the observed performance gap between two-set and three-set splitting strategies is statistically stable, rather than an artifact arising from random partition.

\begin{figure}[h!]
\centering
\includegraphics[width=1\textwidth]{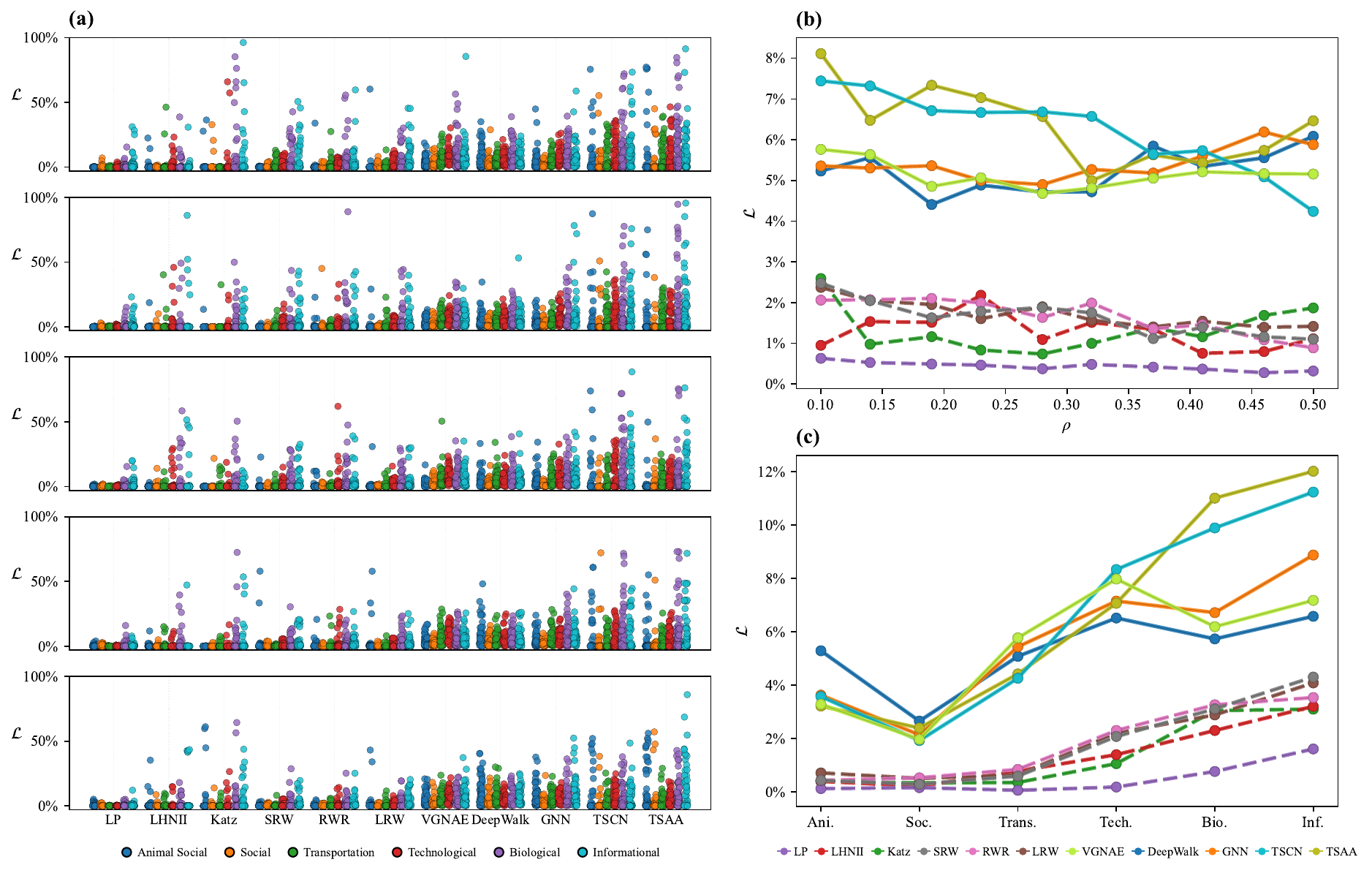}
\caption{
Performance of different link prediction algorithms under varying $\rho$.
\textbf{(a)} Scatter plot of the $\mathcal{L}$ distribution for each algorithm across six domains of networks. The x-axis represents algorithms, and the y-axis represents $\mathcal{L}$. Each colored dot corresponds to a different network category.
\textbf{(b)} The average $\mathcal{L}$ of different algorithms as the training ratio $\rho$ varies.
\textbf{(c)} The average $\mathcal{L}$ of each algorithm across different network categories, where the x-axis denotes network types and the y-axis denotes $\mathcal{L}$.
In panels (b) and (c), dashed lines represent heuristic and random-walk-based methods, while solid lines represent deep learning methods and similarity-transfer methods.
}
\label{lossratio}
\end{figure}

Figure~\ref{lossratio}(a) presents the distribution of $\mathcal{L}$ for different link prediction algorithms across six categories of networks under varying training ratios $\rho$. Collectively, heuristic and random-walk-based algorithms exhibit consistently low $\mathcal{L}$ values with limited variance across different network types and $\rho$ settings. In contrast, deep learning and similarity-transfer methods show significantly higher $\mathcal{L}$ and larger fluctuations, particularly on Informational networks. Figure~\ref{lossratio}(b) shows the average $\mathcal{L}$ of each algorithm at different training ratios $\rho$. It can be observed that deep learning and similarity-transfer methods suffer from higher loss rates, approximately 5–8\%, whereas heuristic and random-walk-based methods display generally flat and stable curves with relatively small fluctuations, indicating that these methods can maintain robust performance across varying levels of training data availability. Figure~\ref{lossratio}(c) compares the transfer performance of algorithms across different network categories. The results reveal that deep learning and similarity-transfer methods exhibit much larger performance disparities across networks. In particular, the $\mathcal{L}$ of the TSAA algorithm increases from approximately 2\% on Social networks to around 12\% on Informational networks. Overall, the two classes of algorithms form a clear stratification in transfer stability: heuristic and random-walk-based methods remain at low loss levels with narrow variance, while deep learning and similarity-transfer methods demonstrate higher and more volatile loss rates. In general, heuristic and random-walk-based algorithms show greater robustness and consistency in hyperparameter migration, which is consistent with the statistical results of Table \ref{tab:lossratio}.

\begin{figure}[h!]
\centering
\includegraphics[width=1\textwidth]{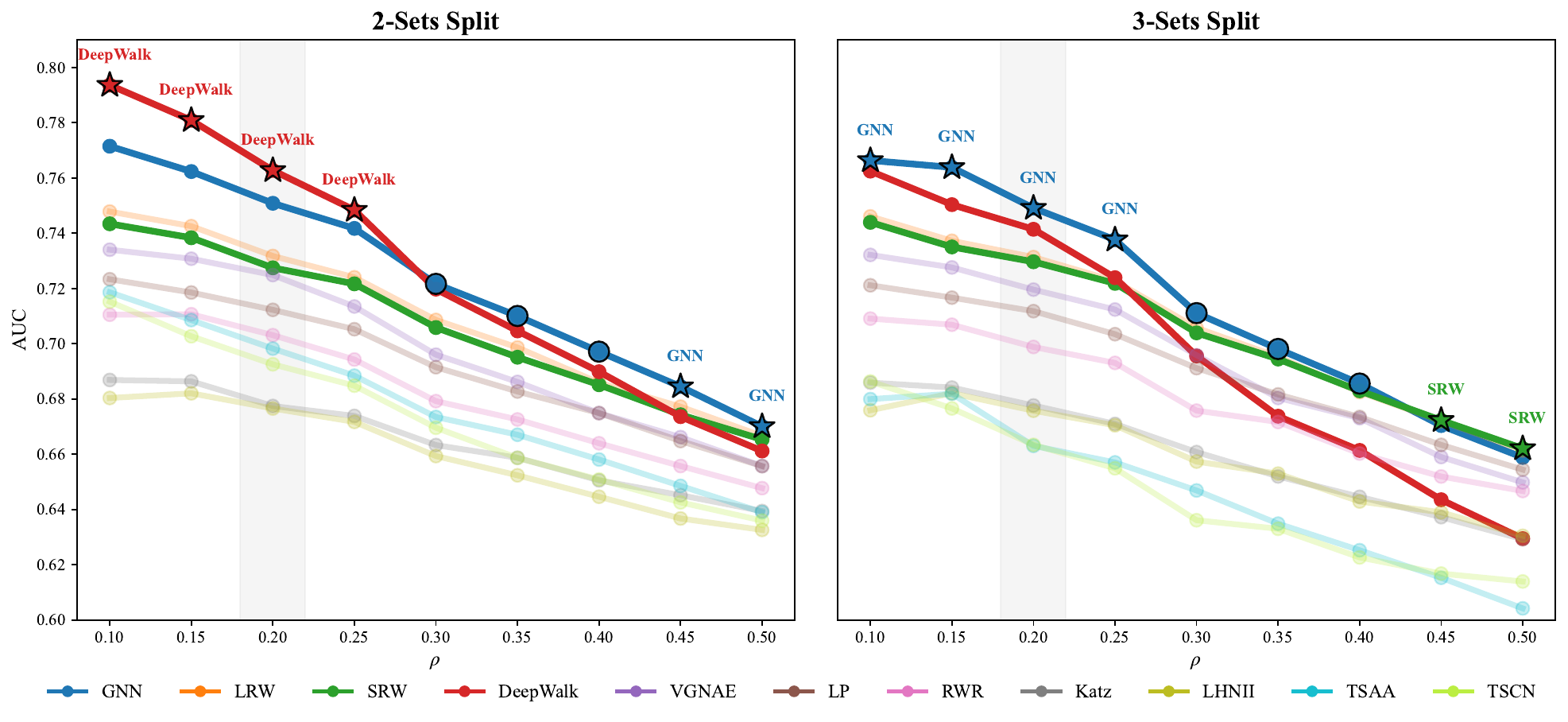}
\caption{AUC under different training ratios $\rho$. The left and right panels correspond to the two-set split and three-set split settings, respectively. Star markers denote algorithms that achieve the best performance under a specific $\rho$ in only one of the two split settings, while enlarged circular markers indicate algorithms that are optimal under both two-set split and three-set split for the same $\rho$. The grey vertical band highlights $\rho = 0.2$, which matches the setting adopted in Table~\ref{tab:lossratio}.}
\label{AUC}
\end{figure}

The training ratio $\rho$ in experiments ranges from $0.1$ to $0.5$. In general, the AUC values under the two-set split are consistently higher than those under the three-set split. This observation is intuitive, as the two-set split implicitly leverages information from the test set during partitioning, leading to an overestimation of model performance. Figure~\ref{AUC} illustrates the AUC variation with respect to $\rho$ under both data splittling strategies. GNN, LRW, and SRW exhibit strong and stable performance across all $\rho$ values in both splits, indicating that these algorithms possess high robustness under varying training conditions. In contrast, the performance of DeepWalk, TSAA, and TSCN is more sensitive. These algorithms not only achieve substantially higher AUC under the two-set split compared with the three-set split, but also rank noticeably better in the two-set setting, suggesting a significant performance drop when the available training data become limited.

Further more, the comparison reveals that the performance trends of different algorithms with respect to $\rho$ vary considerably. In general, as $\rho$ increases, the AUC tends to decrease, which is attributed to the reduction in the number of available training samples, i.e., $(1-\rho)^2$. Notice that, not only is there a consistent performance gap between the two split settings, but the relative ranking of algorithms also changes substantially. For example, when $\rho \in [0.1,0.25]$, DeepWalk achieves the best performance under the two-set split, while under the three-set split it is surpassed by GNN. This demonstrates that the relative advantages of algorithms can shift depending on the chosen data splitting strategy. More seriously, when $\rho$ takes a large value, the actual deserved performance of DeepWalk (based on the three-data split) is already among the three worst-performing ones; however, if the two-set split is adopted, it remains among the top two or three algorithms. Such discrepancies indicate that the superior performance of certain methods under the two-set split is partially attributable to information leakage, whereas the more rigorous three-set split reveals their true effectiveness.

\section{Discussion}

This study compares the differences between two-set split and three-set split in link prediction tasks, and introduces the Loss Ratio $\mathcal{L}$ as a metric to quantify the bias caused by information leakage under two-set split. Figure~\ref{AUC} shows that, under identical network and sampling conditions, the two-set split consistently yields higher AUC values, whereas the three-set split, which trains hyperparameters on the validation set, has lower performance. The AUC–$\lambda$ curves under the two splitting strategies enable direct quantification of this discrepancy. A larger $\mathcal{L}$ indicates that two-set split substantially overestimates performance on unseen data. This bias occurs because two-set split uses the same partition for both model selection and evaluation, causing the hyperparameters to overfit the evaluation data. In contrast, three-set split employs an independent validation set for hyperparameter tuning, thereby preventing test information from leaking and yielding a more realistic estimate of model generalization. To ensure strict comparability on the same test set, we further adopt a nested splitting strategy, where the validation set is drawn from the training portion of the two-set split and transfer evaluation is conducted on the shared test set.

Heuristic methods and random-walk-based methods have higher robustness across multiple networks. The results from these two approaches exhibited lower and more concentrated mean loss ratios compared to neural network methods, demonstrating reduced sensitivity to validation schemes and sample size. In particular, algorithms such as LRW, SRW, and LP achieved a balance between stability and high performance. We believe this is because they involve fewer parameters and are less dependent on validation sets than neural network algorithms. In contrast, similarity-transfer algorithms and deep learning methods exhibited higher and more dispersed loss ratios. This suggests that their results are more susceptible to variations in sample size and partitioning, implying that their high performance is more sensitive to training data size and parameter tuning strategies.

From the network perspective, Informational and Biological networks exhibit consistently higher $\mathcal{L}$ values, suggesting that performance fluctuations are more likely to occur under different splits and that model generalization on such networks is inherently more challenging. Researchers should therefore exercise greater caution when evaluating models on these networks, as using two-set split may lead to overly optimistic conclusions. In contrast, Social networks tend to show lower $\mathcal{L}$, indicating that their structural patterns are more stable under different splits, resulting in smaller performance variations.

It is also worth noting that the training ratio $\rho$ significantly affects the absolute performance of algorithms. At different values of $\rho$, the relative ranking of algorithms under two-set split and three-set split may change. For example, when $\rho = 0.2$, DeepWalk achieves the highest AUC under two-set split, yet fails to maintain its advantage under three-set split. Such rank inversion reflects the sensitivity of parameterized models to splitting strategies and highlights the risk that using the evaluations from two-set split may obscure the true capabilities of an algorithm.

In summary, the apparent superiority of certain prediction models may stem from inappropriate evaluation protocols rather than intrinsic algorithmic advantages. The three-set split should be adopted as the standard evaluation protocol, particularly for network categories where models are more prone to overfitting. Heuristic and random-walk-based methods offer more robust performance when data scale or parameter search is restricted, while deep learning methods should be paired with an independent validation set and carefully controlled hyperparameter space to avoid overstated results. This study provides a novel perspective that suggests to re-evaluate link prediction algorithms. The findings not only offer practical guidance for model deployment but also establish an empirical foundation for developing fairer and more reproducible evaluation paradigms in the field.

\section*{Acknowledgments}
More than a decade ago, one of the authors of this paper (Tao Zhou) had an argument with Dr. Junming Huang regarding whether the two-set split or the three-set split should be used in link prediction. Caused by naivety and a lack of experience in artificial intelligence, Tao Zhou believed that the two-set split was feasible, which was also the mainstream method at that time. They co-authored a paper titled "Predicting missing links via significant paths" [EPL 106 (2014) 18008] using the two-set splitting strategy. Notably, the parameterized algorithm used as a benchmark in that Letter was the local path index, whose results were surprisingly close in the two data splits (as shown in this paper). Due to their differing viewpoints, Dr. Junming Huang later withdrew his name as a co-author, as indicated in the erratum [Erratum: Predicting missing links via significant paths, EPL 108 (2014) 49901].
The results of this paper strongly demonstrate that Dr. Junming Huang was correct! This work was partially supported by the National Natural Science Foundation of China (Nos. T2293771 and 42361144718) and the STI 2030–Major Projects (No. 2024ZD0523903). Y.L. acknowledges financial support from the University of Warwick and the China Scholarship Council (CSC) under the Scholarship Fund (File No. 202508060219).

\end{CJK}
\end{document}